\documentclass[twocolumn]{aastex631}
\usepackage{newtxtext,newtxmath}

\usepackage[T1]{fontenc}

\usepackage{graphicx}	
\usepackage{amsmath}	
\usepackage{float}
\usepackage{ulem} 
\usepackage[autopunct=true]{csquotes}
\newcommand\soutpars[1]{\let\helpcmd\sout\parhelp#1\par\relax\relax} 
\long\def\parhelp#1\par#2\relax{
  \helpcmd{#1}\ifx\relax#2\else\par\parhelp#2\relax\fi
} 

\newcommand{\datastatement}[1]{\begin{small}\section*{Data Availability Statement}\end{small}{\noindent #1}\vspace{5pt}}

\newcommand{\Dt}[1]{\frac{\mathrm{d} #1}{\mathrm{dt}}}
\newcommand{\initvalupper}[1]{#1^{0}}

\newcommand{\driftvel}{{\bf w}_{s}}
\newcommand{\driftvelmag}{w_{s}}
\newcommand{\dustvel}{{\bf v}_{d}}
\newcommand{\gasvel}{{\bf u}_{g}}
\newcommand{\gasden}{\rho_{g}}

\newcommand{\dustden}{\rho_{d}}

\newcommand{\ts}{t_{s}}
\newcommand{\cs}{c_{s}}

\newcommand{\tL}{t_{L}}
\newcommand{\grainsuff}{_{\rm grain}}
\newcommand{\internaldensity}{\bar{\rho}\grainsuff^{\,i}}
\newcommand{\grainsize}{\epsilon\grainsuff}

\newcommand{\grainsizemax}{\grainsize^{\rm max}}

\newcommand{\B}{{\bf B}}

\newcommand{\Bhat}{\hat\B}

\newcommand{\dustgas}{\mu^{\rm dg}}

\def\app#1#2{%
  \mathrel{%
    \setbox0=\hbox{$#1\sim$}%
    \setbox2=\hbox{%
      \rlap{\hbox{$#1\propto$}}%
      \lower1.1\ht0\box0%
    }%
    \raise0.25\ht2\box2%
  }%
}

\DeclareOldFontCommand{\bf}{\normalfont\bfseries}{\mathbf} 
\providecommand{\DIFdel}[1]{} 
\providecommand{\DIFaddbegin}{} 
\providecommand{\DIFaddend}{} 
\providecommand{\DIFdelbegin}{} 
\providecommand{\DIFdelend}{} 
\providecommand{\DIFaddbeginFL}{} 
\providecommand{\DIFaddendFL}{} 
\providecommand{\DIFdelbeginFL}{} 
\providecommand{\DIFdelendFL}{} 
\newcommand{\DIFscaledelfig}{0.5}
\RequirePackage{settobox} 
\RequirePackage{letltxmacro} 
\newsavebox{\DIFdelgraphicsbox} 
\newlength{\DIFdelgraphicswidth} 
\newlength{\DIFdelgraphicsheight} 
\LetLtxMacro{\DIFOincludegraphics}{\includegraphics} 
\newcommand{\DIFaddincludegraphics}[2][]{{\color{blue}\fbox{\DIFOincludegraphics[#1]{#2}}}} 
\newcommand{\DIFdelincludegraphics}[2][]{
\sbox{\DIFdelgraphicsbox}{\DIFOincludegraphics[#1]{#2}}
\settoboxwidth{\DIFdelgraphicswidth}{\DIFdelgraphicsbox} 
\settoboxtotalheight{\DIFdelgraphicsheight}{\DIFdelgraphicsbox} 
\scalebox{\DIFscaledelfig}{
\parbox[b]{\DIFdelgraphicswidth}{\usebox{\DIFdelgraphicsbox}\\[-\baselineskip] \rule{\DIFdelgraphicswidth}{0em}}\llap{\resizebox{\DIFdelgraphicswidth}{\DIFdelgraphicsheight}{
\setlength{\unitlength}{\DIFdelgraphicswidth}
\begin{picture}(1,1)
\thicklines\linethickness{2pt} 
{\color[rgb]{1,0,0}\put(0,0){\framebox(1,1){}}}
{\color[rgb]{1,0,0}\put(0,0){\line( 1,1){1}}}
{\color[rgb]{1,0,0}\put(0,1){\line(1,-1){1}}}
\end{picture}
}\hspace*{3pt}}} 
} 
\LetLtxMacro{\DIFOaddbegin}{\DIFaddbegin} 
\LetLtxMacro{\DIFOaddend}{\DIFaddend} 
\LetLtxMacro{\DIFOdelbegin}{\DIFdelbegin} 
\LetLtxMacro{\DIFOdelend}{\DIFdelend} 
\DeclareRobustCommand{\DIFaddbegin}{\DIFOaddbegin \let\includegraphics\DIFaddincludegraphics} 
\DeclareRobustCommand{\DIFaddend}{\DIFOaddend \let\includegraphics\DIFOincludegraphics} 
\DeclareRobustCommand{\DIFdelbegin}{\DIFOdelbegin \let\includegraphics\DIFdelincludegraphics} 
\DeclareRobustCommand{\DIFdelend}{\DIFOaddend \let\includegraphics\DIFOincludegraphics} 
\LetLtxMacro{\DIFOaddbeginFL}{\DIFaddbeginFL} 
\LetLtxMacro{\DIFOaddendFL}{\DIFaddendFL} 
\LetLtxMacro{\DIFOdelbeginFL}{\DIFdelbeginFL} 
\LetLtxMacro{\DIFOdelendFL}{\DIFdelendFL} 
\DeclareRobustCommand{\DIFaddbeginFL}{\DIFOaddbeginFL \let\includegraphics\DIFaddincludegraphics} 
\DeclareRobustCommand{\DIFaddendFL}{\DIFOaddendFL \let\includegraphics\DIFOincludegraphics} 
\DeclareRobustCommand{\DIFdelbeginFL}{\DIFOdelbeginFL \let\includegraphics\DIFdelincludegraphics} 
\DeclareRobustCommand{\DIFdelendFL}{\DIFOaddendFL \let\includegraphics\DIFOincludegraphics} 
\RequirePackage{listings} 
\lstdefinelanguage{DIFcode}{ 
  moredelim=[il][\color{white}\tiny]{\%DIF\ <\ }, 
  moredelim=[il][\sffamily\bfseries]{\%DIF\ >\ } 
} 
\lstdefinestyle{DIFverbatimstyle}{ 
	language=DIFcode, 
	basicstyle=\ttfamily, 
	columns=fullflexible, 
	keepspaces=true 
} 
\lstnewenvironment{DIFverbatim}{\lstset{style=DIFverbatimstyle}}{} 
\lstnewenvironment{DIFverbatim*}{\lstset{style=DIFverbatimstyle,showspaces=true}}{} 

\begin{document}

\title[STARFORGE: Dust-Evacuated Zones]{Dust-Evacuated Zones Near Massive Stars: Consequences of Dust Dynamics on Star-forming Regions}
\shorttitle{STARFORGE: Dust-Evacuated Zones}
 \shortauthors{Soliman et al.}

\correspondingauthor{Nadine H.~Soliman}
\email{nsoliman@caltech.edu}
\author[0000-0002-6810-1110]{Nadine H.~Soliman}
\affiliation{TAPIR, Mailcode 350-17, California Institute of Technology, Pasadena, CA 91125, USA}

\author{Philip F.~Hopkins}
\affiliation{TAPIR, Mailcode 350-17, California Institute of Technology, Pasadena, CA 91125, USA}

\author{Michael Y. Grudi\'{c}}
\affiliation{Carnegie Observatories, 813 Santa Barbara St, Pasadena, CA 91101, USA}

\begin{abstract}

Stars form within dense cores composed of both gas and dust within molecular clouds. However, despite the crucial role that dust plays in the star formation process, its dynamics is frequently overlooked, with the common assumption being a constant, spatially uniform dust-to-gas ratio and grain size spectrum. In this study, we introduce a set of radiation-dust-magnetohydrodynamic simulations of star forming molecular clouds from the {\small STARFORGE} project. These simulations expand upon the earlier radiation MHD models, which included cooling, individual star formation, and feedback. Notably, they explicitly address the dynamics of dust grains, considering radiation, drag, and Lorentz forces acting on a diverse size spectrum of live dust grains. We find that once stars exceed a certain mass threshold ($\sim 2 M_{\odot}$), their emitted radiation can evacuate dust grains from their vicinity, giving rise to a dust-suppressed zone of size $\sim 100$ AU. This removal of dust, which interacts with gas through cooling, chemistry, drag, and radiative transfer, alters the gas properties in the region. Commencing during the early accretion stages and preceding the Main-sequence phase, this process results in a mass-dependent depletion in the accreted dust-to-gas (ADG) mass ratio within both the circumstellar disc and the star. We predict massive stars ($\gtrsim 10 M_{\odot}$) would exhibit ADG ratios that are approximately one order of magnitude lower than that of their parent clouds. Consequently, stars, their discs, and circumstellar environments would display notable deviations in the abundances of elements commonly associated with dust grains, such as carbon and oxygen.

\end{abstract}



\section{Introduction}

\label{sec:intro}

Star formation and stellar evolution are complex processes influenced by a multitude of physical mechanisms and environmental factors within Giant Molecular Clouds (GMCs) \citep{ McKee2007, Girichidis2020}. The presence of magnetized, supersonic, and turbulent flows within these clouds drives strong density fluctuations, giving rise to regions with varying densities and spatial dimensions \citep{Larson1981,MacLow2004}. As these density fluctuations develop, some reach a critical point where their gravitational force begins to dominate, initiating protostellar collapse. The collapse of these gas and dust over-densities gives birth to protostars, which accrete nearby material and evolve into mature stars. During this process, they dynamically interact with the surrounding cloud through feedback mechanisms such as radiation, jets, radiatively-driven stellar winds, and supernova explosions \citep{Krause2020}.

Dust, an essential component of GMCs, heavily influences all of these processes. These particles, known as ``grains'', are created as a byproduct of stellar evolution, within the atmospheres of evolved stars and in supernova remnants. Dust grains reprocess stellar radiation, absorbing and scattering far-ultraviolet (FUV), optical and infrared (IR) photons, re-emitting them in the IR and submillimeter wavelengths \citep{draine1984optical, mathis1990interstellar, li2001infrared,tielens:2005.book}. Furthermore, they scatter background radiation and emit a thermal continuum in the IR.  In addition to its radiative interactions, dust governs the thermodynamics and chemistry of gas within the ISM, exerting a substantial influence on the intricate processes triggering star formation 
\citep{whittet1993dust, salpeter1977formation, weingartner2001electron,weingartner:2001.grain.charging.photoelectric,draine:2003.dust.review, dorschner:dust.mineralogy.review,  spitzer2008physical, watanabe2008ice, minissale2016dust}. Moreover, dust acts as a reservoir, confining heavy elements within a solid phase, which can subsequently become integrated into stars and planets, thereby significantly influencing their overall compositions
\citep{whittet1993dust, salpeter1977formation,  dorschner:dust.mineralogy.review,  spitzer2008physical, watanabe2008ice, minissale2016dust}. Dust regulates the temperature of the gas through photoelectric heating, collisional heating and cooling, and efficient radiative cooling \citep{weingartner2001electron,weingartner:2001.grain.charging.photoelectric,draine:2003.dust.review}. Moreover, dust serves as a reservoir for metals, which deplete onto the dust grains. These metals can later become integrated into forming stars and planets, influencing their overall compositions . Once luminous sources form within the GMC, the presence of dust can also initiate radiatively driven outflows in the surrounding regions \citep{ murray2005maximum,king2015powerful, 2018A&ARv..26....1H}.

 In addition, the dynamic interplay between dust and gas is pivotal in shaping the evolution of star-forming regions. Turbulence within the cloud can generate variations in the dust density on scales comparable to the turbulent eddy turnover scale, leading to substantial fluctuations in the dust-to-gas (DTG) ratio within prestellar cores \citep{thoraval:1997.sub.0pt04pc.no.cloud.extinction.fluct.but.are.on.larger.scales, thoraval:1999.small.scale.dust.to.gas.density.fluctuations, abergel:2002.size.segregation.effects.seen.in.orion.small.dust.abundances, flagey2009evidence,boogert2013infrared}. As demonstrated by \citet{hopkins:totally.metal.stars}, this phenomenon has significant implications within large GMCs, where dust dynamics alone could potentially lead to stellar populations that exhibit notable variations in abundances of elements commonly found in dust grains, including CNO, Si, Mg and Fe \citep{hopkins.conroy.2015:metal.poor.star.abundances.dust}.

Despite its well-recognized importance, most simulations that model star-formation often assume a perfect coupling between the dust and the surrounding gas. This treatment involves considering both components as moving together, with dust essentially acting as an ``additional opacity" to the gas. However, the dynamics of dust is far more complex. Dust particles, ranging from a few angstroms to several micrometres in size, are inherently aerodynamic and often charged. When dust grains are accelerated by forces like radiation pressure, they drift through the gas, and the imparted momentum couples to the surrounding gas through electromagnetic and drag forces, redistributing momentum in the environment. The efficiency of this coupling relies on both the characteristics of the grains and the surrounding environment. For typical densities of GMCs, where $n_{\rm H} \sim 10^2 \, \rm cm^{-3}$, the collisional interactions between dust and gas are relatively infrequent, resulting in a relatively weak coupling of the dust dynamics with that of the gas. This decoupling between grain and gas populations has notably been observed, particularly in the case of larger dust grains \citep{kruger:2001.ulysses.large.grain.and.pioneer.compilation.data, frisch:2003.small.large.grains.decoupled.near.solar.neighborhood, meisel:2002.radar.micron.dust.ism.particles.many.large.grains, altobelli:2006.helios.large.local.interstellar.grains,altobelli:2007.cassini.confirms.large.microns.sized.interstellar.dust.grains, poppe:2010.new.horizons.confirms.ulysses.large.dust.measurements}. To improve our understanding of dust dynamics within this environment, this study serves as an initial investigation that relaxes some assumptions regarding the perfect coupling of gas and dust dynamics.

In this work, we present the first radiation-dust-magnetohydrodynamic ({\small{RDMHD}}) simulations of star-forming molecular clouds that explicitly account for dust dynamics. We utilize simulations from the STAR FORmation in Gaseous Environments ({\small STARFORGE}) project, which provide a comprehensive representation of individual star formation. This encompasses the stages of proto-stellar collapse, subsequent accretion and stellar feedback, main-sequence evolution, and stellar dynamics, with a thorough consideration of the relevant physical processes (as described in \citet{starforge.fullphysics}). Within the context of this paper, our focus is directed towards investigating the impacts of explicit dust-gas radiation dynamics on the properties of the stellar populations that emerge within the cloud.

The paper is structured as follows: In Section \ref{sec:2}, a concise outline of the code is presented (for a detailed description of the numerical methods and implementations, refer to \citet{grudic:starforge.methods}) along with a description of the initial conditions (ICs) of the simulations. We discuss the results obtained from our fiducial simulation runs in Section \ref{sec:res} and compare them to runs with simplified dust physics. Finally, we discuss the implications of our findings in Section \ref{sec:dis} and present our conclusions in Section \ref{sec:conc}.

\section{Simulations}
\label{sec:2}
\subsection{STARFORGE Physics}
We run a set of RDMHD {\small STARFORGE} simulations to study star formation in GMCs, using the {\small GIZMO} code \citep{hopkins:gizmo}. These simulations adopt the complete physics setup from the ``full'' {\small STARFORGE} model, as described in \citet{starforge.fullphysics}. For solving the equations of ideal magnetohydrodynamics (MHD), we rely on the {\small GIZMO} Meshless Finite Mass MHD solver, detailed in \citet{hopkins:gizmo.mhd} and \citet{hopkins:gizmo.mhd.cg}. In this study, we extend the standard {\small STARFORGE} physics by explicitly modeling the dynamics of the dust particles, where the details of the implementation are outlined in Section \ref{sec:dust}.

We utilize the GIZMO meshless frequency-integrated M1 solver \citep{lupi:2017.gizmo.galaxy.form.methods, lupi:2018.h2.sfr.rhd.gizmo.methods, hopkins.grudic:2018.rp, hopkins:radiation.methods, grudic:starforge.methods, hopkins.grudic:2018.rp, hopkins:radiation.methods} to evolve the time-dependent, frequency-integrated radiative transfer (RT) equation adopting a reduced speed of light $c = 30 \, \rm km s^{-1}$. Adopting a reduced speed of light is a common practice in star formation simulations, as it allows for larger time-steps while ensuring the reduced speed still exceeds other relevant speeds in the problem, thereby maintaining the accuracy of radiative transfer processes without compromising computational efficiency \citep{geen_2015_rt}. The radiation is discretised into five distinct frequency bands, which cover a range extending from the Lyman continuum to the far infrared (FIR). As radiation interacts with matter, it triggers processes such as photoionisation, photodissociation, photoelectric heating, and dust absorption. 
Additionally, the matter undergoes radiative cooling and heating, accounting for metal lines, molecular lines, fine structure lines, as well as continuum and dust collisional processes, as outlined in \citet{fire3}.
An especially noteworthy feature of our radiative transfer approach is the direct coupling of each radiation band to the dust particles. A more comprehensive description of this integration is presented in Section \ref{sec:dust}. 
The radiation field is initialised with an external heating source at the boundaries, representing the interstellar radiation field (ISRF). As local sources (stars) emerge, they also contribute to the radiation field.

Individual stars in the simulations are represented by sink particles. These sink particles arise from gas cells that meet the criteria for runaway gravitational collapse and follow the protostellar evolution model introduced by \citet{offner_protostar_mf}. Sink particles can then accrete bound gas and dust elements, where dust accretion adheres to the same critera as gas accretion. Each sink particle separately tracks the quantities of both dust and gas it has accumulated since its initial sink formation. To ensure realistic accretion irrespective of resolution, the sink particles first accrete particles onto an unresolved disk reservoir, the material in which is then smoothly accreted onto the sink particle. 

The sink particles undergo growth through accretion while progressing along the main sequence. Their luminosity and radius follow the relationships outlined in \citet{tout_1996_mass_lum}, and they emit a black-body spectrum with an effective temperature of $T_{\rm eff} = 5780,\rm K \left(L_\star/R_\star^2\right)^{1/4}$. Beyond radiation, sinks interact with their surrounding medium via protostellar jets, stellar winds, and the potential occurrence of supernovae. To calculate gravitational forces, we employ an adaptive gravitational softening approach, which spans a range extending down to approximately $ \sim 2 \times10^{-5}$ AU for the gas cells. Additionally, a fixed softening of $6$ AU is utilized for the dust and sink particles. For a more comprehensive description of the modelled physics and numerical techniques, readers are encouraged to refer to \citet{grudic:starforge.methods,starforge.fullphysics}.

\subsection{Dust Physics}
\label{sec:dust}
A detailed description of the numerical methods used for dust modelling in our simulations, albeit focussing on more idealised scenarios, can be found in \citet{hopkins.2016:dust.gas.molecular.cloud.dynamics.sims,lee:dynamics.charged.dust.gmcs,moseley:2018.acoustic.rdi.sims, soliman2022dust}. Following a Monte Carlo sampling approach, we depict dust grains in our simulation as "super-particles" \citep{carballido:2008.grain.streaming.instab.sims,johansen:2009.particle.clumping.metallicity.dependence,bai:2010.grain.streaming.vs.diskparams,pan:2011.grain.clustering.midstokes.sims,2018MNRAS.478.2851M}. Each simulated ``dust particle'' encapsulates an ensemble of dust grains characterized by similar attributes such as grain size $(\epsilon_{\text{grain}})$, mass  $(m_{\text{grain}})$, and charge $(q_{\text{grain}})$ determined through live collisional, photoelectric, and cosmic ray charging \citep{draine:1987.grain.charging,tielens:2005.book}.

The motion of each dust grain is governed by the following equation:

\begin{align}\label{eq:eom}
\Dt{\dustvel} &={\bf a}_{\rm gas,\,dust} + {\bf a}_{\rm grav} + {\bf a}_{\rm rad} \\
\nonumber &=  -\frac{\driftvel}{\ts} - \frac{\driftvel\times\Bhat}{\tL} + {\bf g} + \frac{\pi\,\grainsize^{2}}{m_{\rm grain}\,c}\,\langle Q \rangle_{\rm ext}\,{\bf G}_{\rm rad},
\end{align}

where $\dustvel$ represents the velocity of the grain; ${\bf a}_{\rm gas,\,dust} = -{\driftvel}/{\ts} - {\driftvel\times\Bhat}/{\tL}$ takes into account the forces exerted by the gas on the dust, including drag (quantified by the "stopping time" $\ts$) and Lorentz forces (characterised by the gyro/Larmor time $\tL \equiv m_{\rm grain}c / |q_{\rm grain},{\bf B}|$); $\driftvel \equiv \dustvel - \gasvel$ corresponds to the drift velocity of a dust grain relative to the gas velocity $\gasvel$ at the same position ${\bf x}$; ${\bf B}$ denotes the local magnetic field;  ${\bf a}_{\rm grav} = {\bf g}$ is the gravitational force due to a local gravitational field; and ${\bf a}_{\rm rad}$ is the force due to an incident radiation field ${\bf G}_{\rm rad} \equiv {\bf F}_{\rm rad} - \dustvel \cdot (e_{\rm rad}\,\mathbb{I} + \mathbb{P}_{\rm rad})$ in terms of the radiation flux/energy density/pressure density ${\bf F}_{\rm rad}$, $e_{\rm rad}$, $\mathbb{P}_{\rm rad}$ for a grain of size $\grainsize$ and mass $m_{\rm grain}\equiv (4\pi/3)\,\internaldensity\,\grainsize^{3}$, where $\internaldensity$ is the internal grain mass density, and dimensionless absorption+scattering efficiency $\langle Q \rangle_{\rm ext}$; $c$ is the speed of light. 

Our radiation treatment closely follows the methodology described in \citet{hopkins:2021.dusty.winds.gmcs.rdis, soliman2022dust}, which employed a gray-band opacity assumption. In our current study, we build upon this approach by introducing a five-band opacity treatment for each radiation band evolved by the RT solver. We determine an effective opacity, expressed as $\langle Q \rangle_{\rm ext}(\grainsize, \lambda_{\rm eff}) = \min \left( 2 \pi \grainsize/\lambda_{\rm eff}, 1 \right)$, where $\lambda_{\rm eff} \equiv \sqrt{\lambda_{\min} \lambda_{\max}}$, representing the geometric mean of the wavelength boundaries for the respective radiation bands. For the purposes of calculating the different coefficients within each narrow band, we assume that the opacities are locally gray.

The radiation pressure force on the dust is determined through the M1 radiative transfer method described above, encompassing terms up to $\mathcal{O}(v^{2}/c^{2})$: $\partial_{t} e_{\rm rad} + \nabla \cdot {\bf F}_{\rm rad} = -R_{\rm dust}\,\dustvel \cdot {\bf G}_{\rm rad}/c^{2}$, $\partial_{t} {\bf F}_{\rm rad} + c^{2}\,\nabla \cdot \mathbb{P}_{\rm rad} = -R_{\rm dust}\,{\bf G}_{\rm rad}$. Specifically, it involves equations that govern the temporal evolution of quantities related to the radiation field, where the absorption and scattering coefficients, denoted as $R_{\rm dust}$, are calculated locally from the dust grain distribution.

However, resolving the photon mean free path in global simulations is not always feasible \citep{hopkins:2019.grudic.photon.momentum.rad.pressure.coupling, krumholz:2018.rp}. To address this limitation, we apply a correction factor as outlined by \citet{grudic:starforge.methods} to estimate radiation flux. Photons absorbed by the dust are reprocessed and re-emitted as IR radiation. This re-emitted IR radiation, along with incident IR radiation, undergoes multiple scattering in the medium.

To examine the effects of this implementation, we conduct a series of numerical experiments, systematically introducing and removing the correction factors for each radiation band. Our analysis reveals that while these adjustments can influence the results quantitatively, their effects generally remain within the interquartile range observed in this study.

For all forces originating from the interaction between gas and dust, denoted $a_{\rm gas,dust}$, an equal and opposite force acts on the gas (referred to as ``back-reaction''). For the physical conditions under consideration, the drag experienced by the dust is modeled using the Epstein drag formulation, expressed as:
\begin{align} \label{eq:ts}
\ts &\equiv \sqrt{\frac{\pi \gamma}{8}}\frac{\internaldensity \,\grainsize}{\gasden\,\cs}\, \bigg( 1+\frac{9\pi\gamma}{128} \frac{|\driftvel|^{2}}{\cs^{2}} \bigg)^{-1/2},
\end{align}

where $\gamma$ is the adiabatic index, $\gasden$ is the gas density, and $\cs$ is the local sound speed. Additionally, Coulomb drag is considered, though it typically constitutes a minor correction.

Furthermore, the thermochemistry of the medium is influenced by the presence of dust through both indirect effects, arising from extinction by grains, and direct effects, through dust heating and cooling terms (refer to \cite{starforge.fullphysics} for comprehensive details). These effects depend on the DTG ratio and/or mean grain size. Instead of assuming fixed values, we estimate these parameters by interpolating the local distribution of dust grains to the gas cells, ensuring a self-consistent consideration of the thermodynamics involved. We examine the implications of our explicit dust treatment on the thermodynamics of star-forming regions and delve into the specific influence of varying dust properties, such as grain size, in more detail in \citet{soliman2024thermodynamics}.

The simulations at hand do not resolve the detailed physics of the generation and collimation of protostellar jets. Instead, jets are introduced by spawning high velocity gas cells within a narrow cone (see \citet{starforge.fullphysics} for implementation details). The cone's radius encompasses a few gas cells within the sink radius, and is thus poorly resolved. This may lead to a resolution dependent overestimation of dust ejection by jet particles. To mitigate such spurious effects, we temporarily disable the direct interaction between dust particles and newly spawned cells. This involves interpolating the gas properties from non-jet cells only to dust for a specific duration, allowing the jets to escape the poorly resolved sink radius. Nonetheless, the jet elements can still indirectly affect dust evacuation through their coupling with neighboring gas elements that the dust interacts with. Although this approach might result in missing some physical dust ejections, the narrow opening angle of the jet suggests that any underestimation should be relatively small.

 We experimented with the duration of the decoupling, and find that results converge as long as the decoupling persists until the jets can escape the sink radius or become resolved. We emphasize that we adopt this approach as a precautionary measure to avoid overestimating the extent of dust evacuation. Allowing the interaction to proceed would only lead to further dust ejection near sink particles. Consequently, this approach does not qualitatively affect our findings; instead, it strengthens our conclusions.

\subsection{Initial conditions} 

For our fiducial simulations, we consider an initially uniform-density turbulent molecular cloud with a mass of $M_{\rm cloud}=2 \times 10^{3} \rm M_\odot$ and a radius of $R=3\rm pc$, corresponding to a surface density of $\sim 70 \, \rm M_\odot/pc^2$, and a hydrogen number density of $n_{\rm H} \sim \, \rm 700 cm^{-3}$. The cloud is enveloped within a $30\rm pc$ periodic box with an ambient medium with density $\sim 10^{3}$ times lower than that of the cloud. The initial velocity distribution follows Gaussian random field with with an initial virial parameter $\alpha_{\rm turb} = 5 \sigma^2 R/(3 G M_{\rm cloud}) = 2$, where $G$ is the gravitational constant. The initial magnetic field configuration is uniform, and set to establish a mass-to-flux ratio equivalent to 4.2 times the critical value $\left ( 2\pi G^{1/2}\right)^{-1}$ within the cloud
\citep{nakano_magnetic_fields}.

The cloud is initialized with an initially spatially uniform DTG ratio $\initvalupper{\dustden} = \dustgas \initvalupper{\gasden}$, where $\dustgas = 0.01$ reflects the galactic value. For typical gas cells, the mass resolution in our simulations is $\Delta m_{\rm gas} \sim 10^{-3}\rm M_\odot$, while for the grain super-particles, we refine it to $\Delta m_{\rm dust} \sim 2.5 \times 10^{-6}\rm M_\odot$ ($4\times$ higher resolution for the dust). Furthermore, cells associated with protostellar jets and stellar winds have a higher mass resolution of $10^{-4}\rm M_\odot$. In addition, we extend our sample to include a larger cloud configuration with $M_{\rm cloud}=2 \times 10^{4} \rm M_\odot$ and radius $R=10\rm pc$ corresponding to a surface density of $\sim 70 \, \rm M_\odot/pc^2$, and a hydrogen number density of $n_{\rm H} \sim 200 \, \rm cm^{-3}$. The cloud is also enveloped within a 10$\times$ larger box filled with diffuse material. For the larger cloud, we use a coarser mass resolution, scaled down by a factor of 10.

The dust component is initialized with a net zero drift velocity relative to the surrounding gas. The grain sizes follow an empirical Mathis, Rumpl, \& Nordsieck (MRN) power-law distribution with a differential number density $d N_{\rm d} / d \grainsize \propto \grainsize^{-3.5}$ \citep{mathis:1977.grain.sizes}. The distribution spans a dynamic range of $\grainsize^{\rm max} = 100\grainsize^{\rm min}$, with $\grainsize^{\rm max} = 0.1 \rm \mu m$ for our fiducial simulations. We adopt the classic MRN mixture of carbonaceous $(\sim 40\%)$ and silicate $(\sim 60\%)$ composition, assuming a uniform internal density and composition across different grain sizes. This corresponds to grains with a sublimation temperature of approximately 1500 K and an internal density of $\internaldensity \sim 2.25 \, \rm g/cm^3$. We do not model grain growth or destruction; consequently, we adhere to a fixed size distribution, maintaining constant sizes for particles throughout the simulation.

We provide a summary of the initial conditions for all simulations discussed in this work in Table \ref{table:1}.

\begin{table*}
\begin{tabular}{lllllr}
\hline
Name & $M_{\rm cloud}$ [$\rm M_\odot$] & $R_{\rm cloud}$ [pc] & \text{$\grainsizemax$} [$\rm \mu m$] & \DIFdelbeginFL $\Delta m_{\rm gas}$ [$\rm M_\odot$] & Notes\\ 
\hline
  m2e3\_0.1 &  $2 \times 10^3$  & 3 &0.1  & $10^{-2}$&  fiducial run \\
  m2e3\_0.1\_hires &  $2 \times 10^3$  & 3 &0.1  & $10^{-3}$& $\times$ 10 finer resolution\\
  m2e3\_0.1\_noLor &  $2 \times 10^3$  & 3 &0.1  & $10^{-2}$&  no Lorentz Forces on grains\\
  m2e3\_0.1\_norad &  $2 \times 10^3$  & 3 &0.1  & $10^{-2}$&  no radiation pressure forces on grains\\
  m2e3\_0.1\_pass &  $2 \times 10^3$  & 3&0.1  & $10^{-2}$ & grains only feel drag \& do not exert a back-reaction force\\
  m2e3\_0.1\_nofb &  $2 \times 10^3$  & 3&0.1  & $10^{-2}$ & no stellar winds or protostellar jets \\
  m2e3\_1 &  $2 \times 10^3$  & 3&1.0  & $10^{-2}$ & $\times $ 10 larger grains \\
  m2e3\_10 &  $2 \times 10^3$  & 3&10  & $10^{-2}$ & $\times  100$ larger grains \\
  m2e3\_1\_hires &  $2 \times 10^3$  & 3&1.0  & $10^{-3}$ & $\times $ 10 larger grains \& $\times$10 finer resolution\\
  m2e3\_10\_hires &  $2 \times 10^3$  & 3&10  & $10^{-3}$ & $\times  100$ larger grains \& $\times$10 coarser resolution\\
  \hline
    m2e4\_0.1 &  $2 \times 10^4$  & 10&0.1  & $10^{-2}$ &fiducial run \\
    m2e4\_0.1\_hires &  $2 \times 10^4$  & 10&0.1  & $10^{-3}$ & $\times 10$ finer resolution\\
    m2e4\_0.1\_noLor &  $2 \times 10^4$  & 10&0.1  & $10^{-2}$ & no Lorentz forces on grains \\
    m2e4\_0.1\_norad &  $2 \times 10^4$  & 10&0.1  & $10^{-2}$ & no radiation pressure forces on grains\\
    m2e4\_0.1\_pass &  $2 \times 10^4$  & 10&0.1  & $10^{-2}$ & grains only feel drag \& do not exert a back-reaction force \\
  m2e4\_1 &  $2 \times 10^4$  & 10&1.0  & $10^{-2}$& $\times$10 larger grains \\
  m2e4\_10 &  $2 \times 10^4$  & 10&10  &$10^{-2}$ & $\times$100 larger grains \\
  \hline
  \hline
 \end{tabular}

\caption{The initial conditions for the simulations used in this study. The columns include: {(\bf{1})} Simulation name. {(\bf{2})}  Cloud mass $M_{\rm cloud}$. {(\bf{3})} Cloud Radius $R_{\rm cloud}$. {(\bf{4})} Maximum grain size $\grainsizemax$. {(\bf{5})} Mass resolution $\Delta m_{\rm gas}$. {(\bf{6})} Notes indicating the main variations from the fiducial run.}

\label{table:1}
\end{table*}
\section{Results}
\label{sec:res}
\subsection{Cloud Morphology}
\begin{figure*}
    \centering
    \includegraphics[width=0.8\textwidth]{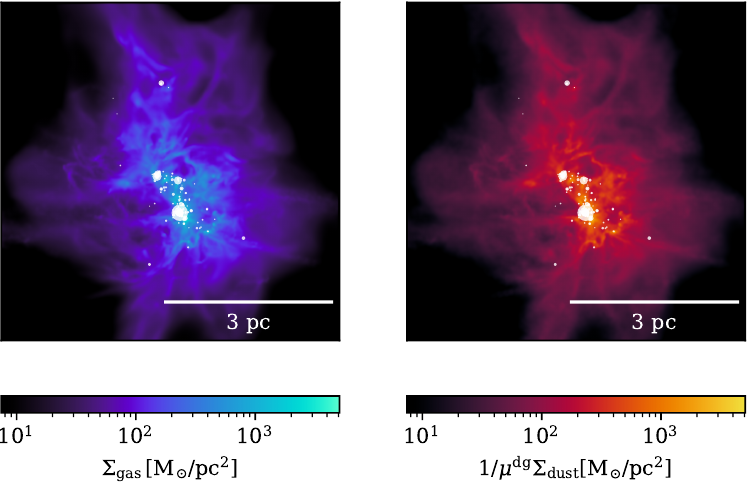}
    \caption{The 2D integrated surface density of gas $\Sigma_{\rm gas}$ (left) and dust $\Sigma-{\rm dust}$ (right) at $t\sim 3.5$ Myrs for the m2e3\_0.1\_hires simulation, corresponding to a cloud with $M_{\rm cloud} \sim 2\times10^3 \rm M_{\rm \odot}$, $\Delta m_{\rm gas} \sim 10^{-3} \rm M_{\rm \odot}$ resolution, and $\grainsizemax = 0.1 \rm \mu m$. This run employs our full physics package and accounts for explicit dust dynamics, including physical phenomena such as dust drag, dust back-reaction, Lorentz forces acting on dust grains, and the explicit computation of dust opacity from grain particles. Note that the dust surface density is scaled by $1/\dustgas$ for ease of comparison. Sink particles, representing stars, are depicted as circles, with their radius reflecting their mass. We note that both the gas and the dust exhibit similar large-scale structural features. For structural differences at smaller scales, refer to Figure \ref{fig:smallscale}. }
    \label{fig:largescale}
\end{figure*}

In Figure \ref{fig:largescale}, we present the 2D integrated surface density of gas, denoted as $\Sigma_{\rm gas}$ displayed on the left and dust, denoted as $\Sigma_{\rm dust}$ shown on the right. These visualizations are from our m2e3\_0.1\_hires simulation, which employs our comprehensive physics model to simulate a cloud with an initial mass of $M_{\rm cloud } \sim 2 \times 10^3 \rm M_{\odot}$ evolved for a duration of $t \sim 3.5$ Myrs.  Sink particles, representing stars in the system, are portrayed as circles with their sizes proportional to their respective masses. The initially spherically uniform cloud undergoes a process of gravitational collapse and fragmentation. This leads to the emergence of a stellar cluster near the central region with a small number of sinks dispersed at greater distances from the cluster's center. When we compare the dust and gas distributions on parsec scales, we observe that the distribution of the dust aligns with that of the gas. This indicates that, on large scales, the dynamics of dust and gas are effectively well-coupled. 

\begin{figure}
    \centering
    \includegraphics[width=0.47\textwidth]{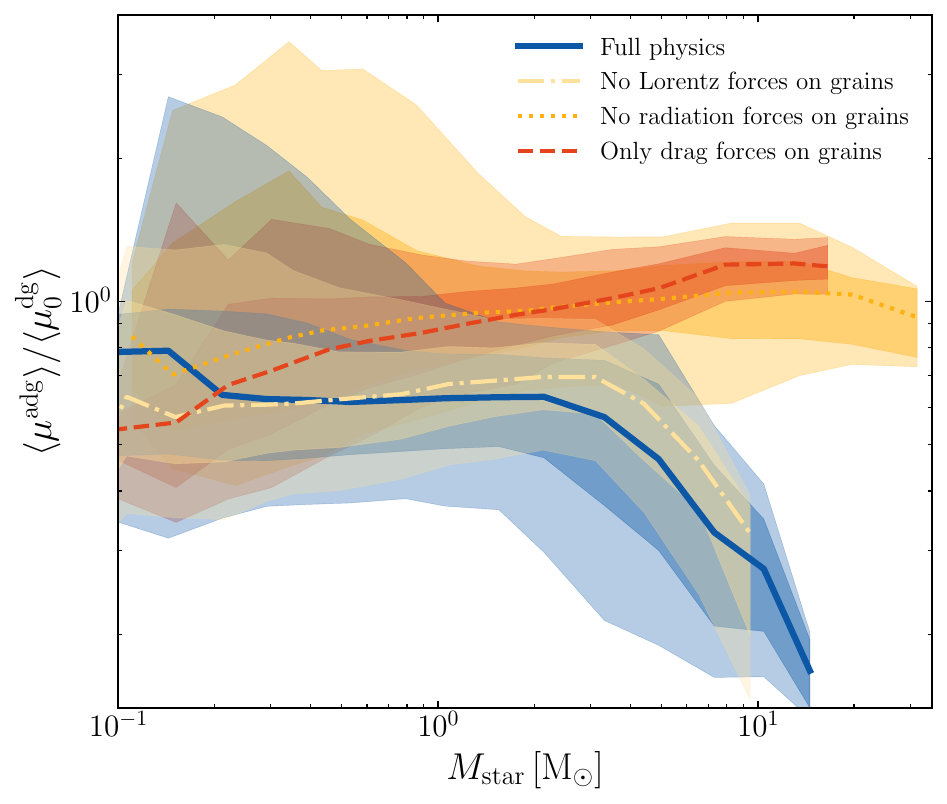}
    \caption{The rolling median of the accreted dust-to-gas (ADG) mass ratio $\mu^{\rm adg}$ calculated for sink particles formed within a cloud of initial mass $M_{\rm cloud} = 2 \times 10^4 \rm M_\odot$. The results are computed within logarithmic mass bins and normalised to the initial mean dust-to-gas (DTG) ratio of the cloud $\langle \mu_0^{\rm dg} \rangle$. The darkly shaded and lightly shaded regions correspond to the interquartile and interdecile ranges respectively. The plot compares the full physics simulation (m2e4\_0.1) to simulations with limited dust physics including: (1) no Lorentz forces act on the grains (m2e4\_0.1\_noLor), (2) no radiation pressure forces act on grains (m2e4\_0.1\_norad), and (3) only aerodynamic drag forces act on grains without back-reaction on the gas (m2e4\_0.1\_pass). Subsolar mass stars exhibit a scattered ADG ratio centered around the average value of $\sim \langle \mu^{\rm dg}_0\rangle$. The fiducial run reveals a clear trend of reduced ADG ratios at higher sink masses influenced by radiation forces on dust, which diminishes when radiative pressure forces are disabled.}
    \label{fig:physics}
\end{figure}

While dust and gas exhibit coherence on cloud-size scales, stars primarily grow by accreting the local mixture of dust and gas, a process that determines various final stellar properties. Consequently, variations in the local DTG ratio can influence the properties of the final stellar population. Due to the simulation's limited spatial resolution for sink accretion ($\sim 10$ AU), the fate of dust, whether it gets accreted by a star or is ejected, remains ambiguous. To address this uncertainty, we calculate a more robust measure: the accreted dust-to-gas (ADG) ratio. This is the ratio of the mass of accreted dust to the mass of accreted gas, i.e., it represents the DTG ratio of the material that composes the star. It is defined as:
\begin{align}
\mu^{\rm adg} & \equiv \frac{M_{\rm accreted, dust}}{M_{\rm accreted, gas}}\approx \frac{M_{\rm accreted, dust}}{M_{\star}} \\
\nonumber& \approx \frac{\sum_{i=1}^{N_{\rm dust}} \Delta m_{\rm dust, i}}{\sum_{i=1}^{N_{\rm gas}} \Delta m_{\rm gas, \rm i} + \sum_{i=1}^{N_{\rm dust}} \Delta m _{\rm dust, i}},
\end{align}
where $M_{\rm accreted, dust}$ is the total mass of the $N_{\rm dust}$ accreted dust elements each of mass $\Delta m_{\rm dust} \sim \dustgas \Delta m_{\rm gas}/4$, and $M_{\rm accreted, gas}$ is the total mass of the $N_{\rm gas}$ accreted gas elements. We acknowledge that this measure has the potential for overestimation, considering the possibility of dust expulsion if accretion were tracked at smaller spatial scales.

\subsection{Reduced Dust Accretion in Massive Stars}

In Figure \ref{fig:physics}, we present the moving median ADG ratio for individual sink particles, calculated within logarithmic stellar mass bins, for a stellar population. We present the sinks formed with our larger cloud with a mass of $M_{\rm cloud} = 2\times 10^4 \rm M_\odot$ to ensure a robust statistical dataset. To discern the contributions of various physical processes to the results, we compare the ADG ratios in our fiducial full-physics run with runs involving limited physics. Specifically, we experiment with runs, including a ``passive'' grain run where grains experience only aerodynamic drag forces, without Lorentz or radiation pressure forces and without exerting any back-reaction forces. We also examine a run in which grains experience both drag and Lorentz forces and, in turn, induce back-reaction forces on the gas.  However, in this specific run, the dust particles neither explicitly experience radiation forces nor is the dust opacity explicitly computed directly from the dust distribution. Instead, the opacity of the gas is calculated by assuming a constant DTG ratio.
 We also consider a scenario where we include all dust physics except Lorentz forces, i.e. the case of neutral dust grains. Finally, we contrast these scenarios with a comprehensive full-physics fiducial run, which encompasses all the aforementioned physical effects, along with explicit coupling of radiation to the dust grains.

 As shown in the figure, we note a larger dispersions in the ADG ratio among sub-solar mass sinks for all simulation runs, with an average value converging to $\sim 1$. This dispersion is primarily ascribed to the inherent limitations in resolution, set at around $\sim 10^{-2} \, \rm M_\odot$, which results in low-mass sinks accreting a relatively limited number of gas and dust elements, and hence exhibiting greater stochastic noise.

For sinks of higher masses ($\geq 2-8 \rm M_\odot$), a trend emerges: the ADG ratio decreases with increasing stellar mass. This phenomenon has implications for stellar properties and the abundance levels of elements commonly found in dust, including CNO, Si, Mg, and Fe, which we anticipate to be significantly reduced. These levels can plummet to nearly an order of magnitude lower than the original solid-phase abundances observed within the cloud. Strikingly, this trend becomes negligible when the dust opacity is not directly computed from the dust distribution, and the grains are not subject to radiation pressure forces. Furthermore, this pattern is virtually indiscernible when the grains are treated as passive entities within the system. We also find that the run without Lorentz forces acting on the grains shows a large statistical overlap with our fiducial physics run within the interquartile range. This implies that Lorentz forces acting on grains do not significantly impact this phenomenon. Hence, we infer that this phenomenon is likely driven by radiation pressure which expels dust grains away from the vicinity of the sink particle. Massive stars, emitting intense radiation, exert substantial radiation pressure on the surrounding dust and gas, thereby significantly affecting their dynamics. This effect is expected to dominate the dynamics when radiation pressure surpasses gravitational forces, a condition expected to be met at specific stellar luminosities and/or masses.

In addition to the aforementioned tests, we conducted simulations to assess the impact of various stellar feedback mechanisms, including protostellar jets, stellar winds, and supernova feedback, by systematically enabling and disabling each in turn. We also explored the influence of magnetic fields through radiative hydrodynamical simulations, distinct from our tests involving neutral grains. We found that the identified trend remains robust across different choices of these model parameters.

 \subsection{Dust Evacuation}
\subsubsection{Toy Model}
We can gain a simple understanding into whether a star would accrete a dust grain by examining the local dust dynamics through a set of simple approximations. Consider a scenario involving an inflowing shell of gas and dust influenced by a combination of drag and radiation from the star pushing the shell outward. The grains should reach a terminal velocity, where they drift relative to the gas at a velocity $\driftvelmag$. For simplicity, let us assume sub-sonic drift with $v_{\rm inflow}\sim \epsilon \cs$ where $\epsilon$ characterises the efficiency of gravitational in-fall with respect to the sound speed, and that the system is in the Rayleigh limit, characterised by $\langle Q\rangle_{\rm ext} \propto \grainsize$, allowing for a closed-form analytic solution. Larger grains with $\langle Q\rangle_{\rm ext} \sim 1$, unlike smaller grains where $\langle Q\rangle_{\rm ext} \propto \grainsize$, would behave similarly to the smaller grains, but would introduce an inverse relation with grain size in Equation \ref{eq:rad}. Note that in this case, assuming isotropic scattering off dust grains, the radiation pressure coefficient $Q_{\rm pr}$, is similar to the extinction coefficient $Q_{\rm pr} \sim \langle Q\rangle_{\rm ext}$. The amount of dust accreted by a star would then depend on the ratio of the terminal drift velocity of dust grains away from the star to their velocity through the in-falling gas $v_{\rm inflow}$. This relationship can be approximated as follows: 

\begin{align}
    \label{eq:rad}
    \frac{w_{\rm s}}{v_{\rm inflow}} &\sim \frac{a_{\rm rad} t_s}{\epsilon c_s} \sim \frac{ \langle Q\rangle _{\rm ext} F_\star }{4 \pi c \epsilon\rho_g  c_s^2}\sim \frac{ \langle Q\rangle_{\rm ext}\,  u_{\rm rad}}{\epsilon u_{\rm thermal}}\\
    \nonumber & \sim \frac{  \langle Q\rangle _{\rm ext} L_{\star}}{4 \pi c r^2 \epsilon \rho_g c_s^2} \\&
    \nonumber \sim  0.1 \left ( \frac{ \langle Q\rangle _{\rm ext} }{0.2}\right) \left (\frac{ L_{\star} }{\rm L_{\odot}} \right )\left (\frac{0.01 \, \rm pc}{r} \right)^2 \left (\frac{1}{\epsilon} \right ) \left (\frac{10^3 \, \rm cm^{-3}}{n_{\rm H}} \right) \left (\frac{20 \rm K}{T} \right ),  
\end{align}

where $F_\star =  L_{\star}/r^2$ is the incident flux for a star of luminosity $ L_{\star}$ on a grain at some radial distance $r$ situated within a region of number density $n_{\rm H}$. Assuming that $L_\star \propto M_\star^{3.5}$, we determine that this ratio reaches unity when $M_\star \sim 2 \rm M_\odot$ given the parameters we consider above.

We emphasize to the readers that this model is intentionally simplistic, designed to offer qualitative physical insight into the relevant parameters for this process. The model does not incorporate detailed models of accretion, density profiles of dust and/or gas, an accurate description of the object's luminosity, or turbulence in the accretion flow, among other factors, all of which are expected to vary spatially and temporally as the star evolves. An additional point worth highlighting is that the ratio does not need to exceed 1 for the ADG ratio to be lower than the DTG ratio of the cloud; even modest values of outward dust drift relative to gas inflow would lead to reduced dust accretion.

\begin{figure*}
    \centering
    \includegraphics[width=0.9\textwidth]{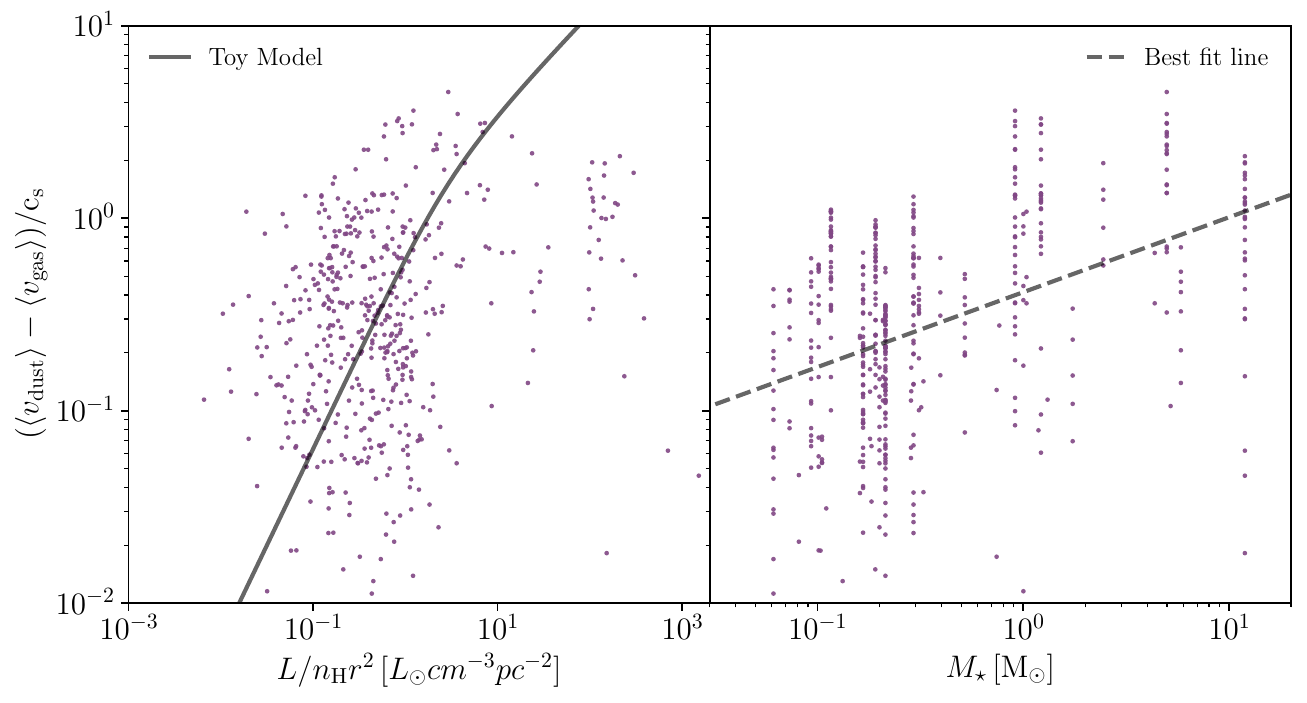}
    \caption{{\bf{Left}}: The mean dust drift velocity $\langle w_s\rangle = \langle v_{\rm dust}\rangle - \langle v_{\rm gas}\rangle$ normalized to the sound speed $c_s$ around sink particles with final $M_\star \gtrsim 2 M_\odot$ across time. Each data point corresponds to mean values measured within a discretized 3D grid centered around the sinks. We plot $\langle v_{\rm dust}\rangle - \langle v_{\rm gas}\rangle$, against the luminosity of the sink particle in its proximity, divided by its gas number density and the square of its radial distance from the sink, $L_\star/n_{\rm H} r^2$. The gray line represents the theoretical prediction from Equation \ref{eq:rad}, incorporating both subsonic $\left ( \langle \driftvelmag \rangle \propto L_\star/n_{\rm H} r^2 \right )$ and supersonic regimes $\left (\langle \driftvelmag \rangle \propto \sqrt{L_\star/n_{\rm H} r^2} \right )$. The dust drift exhibits a positive correlation with $L_\star/n_{\rm H} r^2$, as theoretically predicted. There is scatter away from the theoretical prediction due to the model's simplicity and the influence of other forces and turbulence on the dynamics.
 {\bf{Right}}: The mean drift velocity against the mass of the sink particle. The dashed gray line represents the best fit line}. Note that the discreteness in stellar mass corresponds to the simulation resolution. The dust drift positively correlates with the sink's mass and mean accretion rate, in agreement with predictions.
    \label{fig:v_scatter}
\end{figure*}

We present a test of this hypothesis in Figure \ref{fig:v_scatter}, where we use the full expression for $\ts$ from Equation \ref{eq:ts} to account for the subsonic and supersonic limits. On the left, we show the mean dust drift velocity, calculated as the mean radial velocity of dust relative to a nearby sink subtracted from the mean radial velocity of the gas computed for a discretized 3D grid. This is denoted as $\langle v_{\rm dust}\rangle - \langle v_{\rm gas}\rangle \equiv \driftvelmag$, and the value is then normalized to the local sound speed at the position of the shell. We analyze the relationship between these parameters and the value of $L_\star/n_{\rm H} r^2$ within the grid element.

As expected from the simple relationship described in Equation \ref{eq:rad} plotted in the dashed black line, the dust drift velocity is proportional to $L_\star/n_{\rm H} r^2$, maintaining this proportionality until $\driftvelmag/c_s \sim 1$. Beyond this point, as the flow transitions to the supersonic regime, the correlation transitions to a square root relationship with $L_\star/n_{\rm H} r^2$. We extend this assessment by examining the relationship between mean dust drift velocity and the mass of the central sink particle. Similarly, $\driftvelmag$ positively correlates with the star's mass, or in other words, with the time-integrated accretion rate, which, in turn, is associated with the integrated luminosity of the sink.

\subsubsection{A Case Study}
\begin{figure*}
    \includegraphics[width=0.9\textwidth]{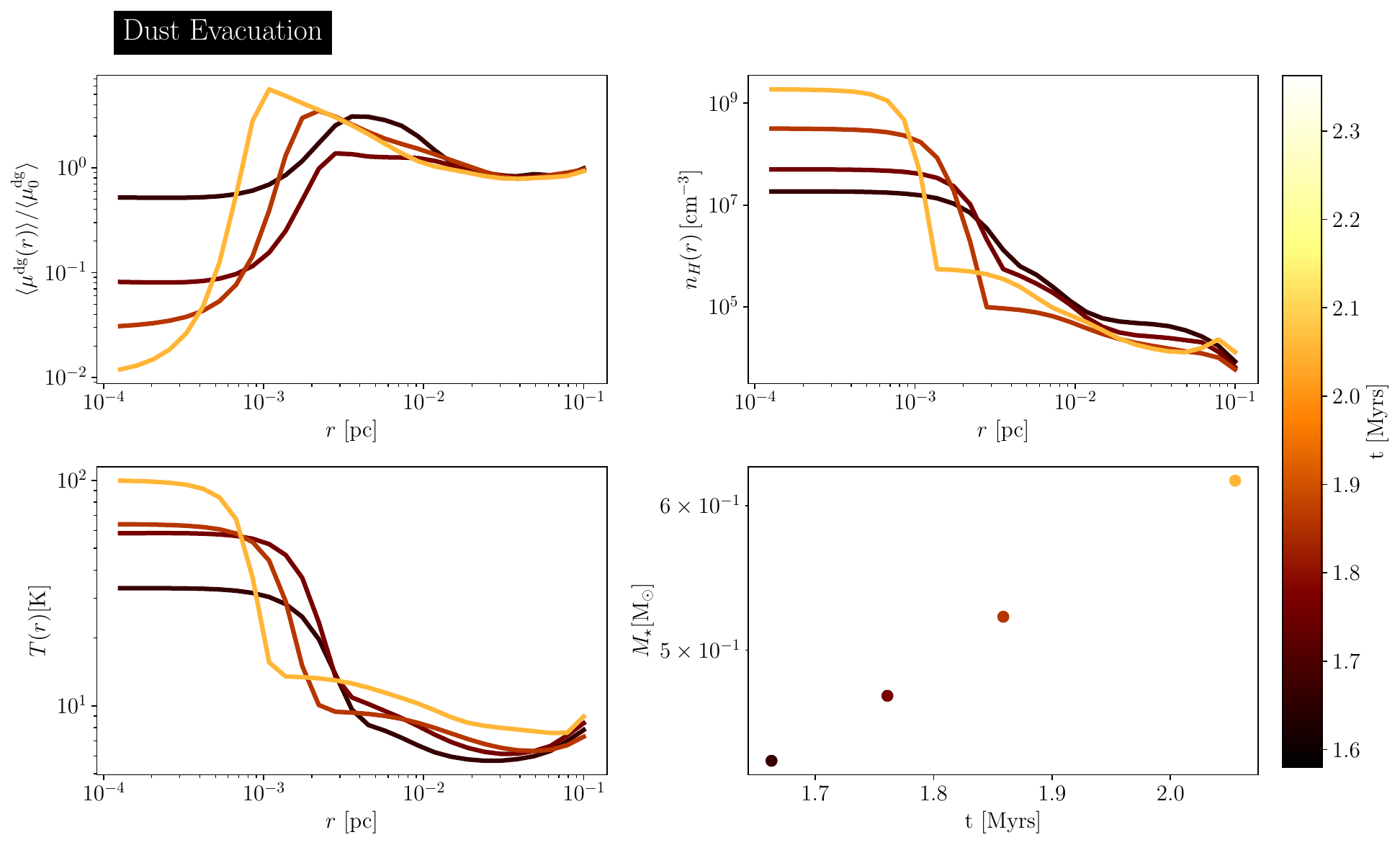}
    \includegraphics[width=0.9\textwidth]{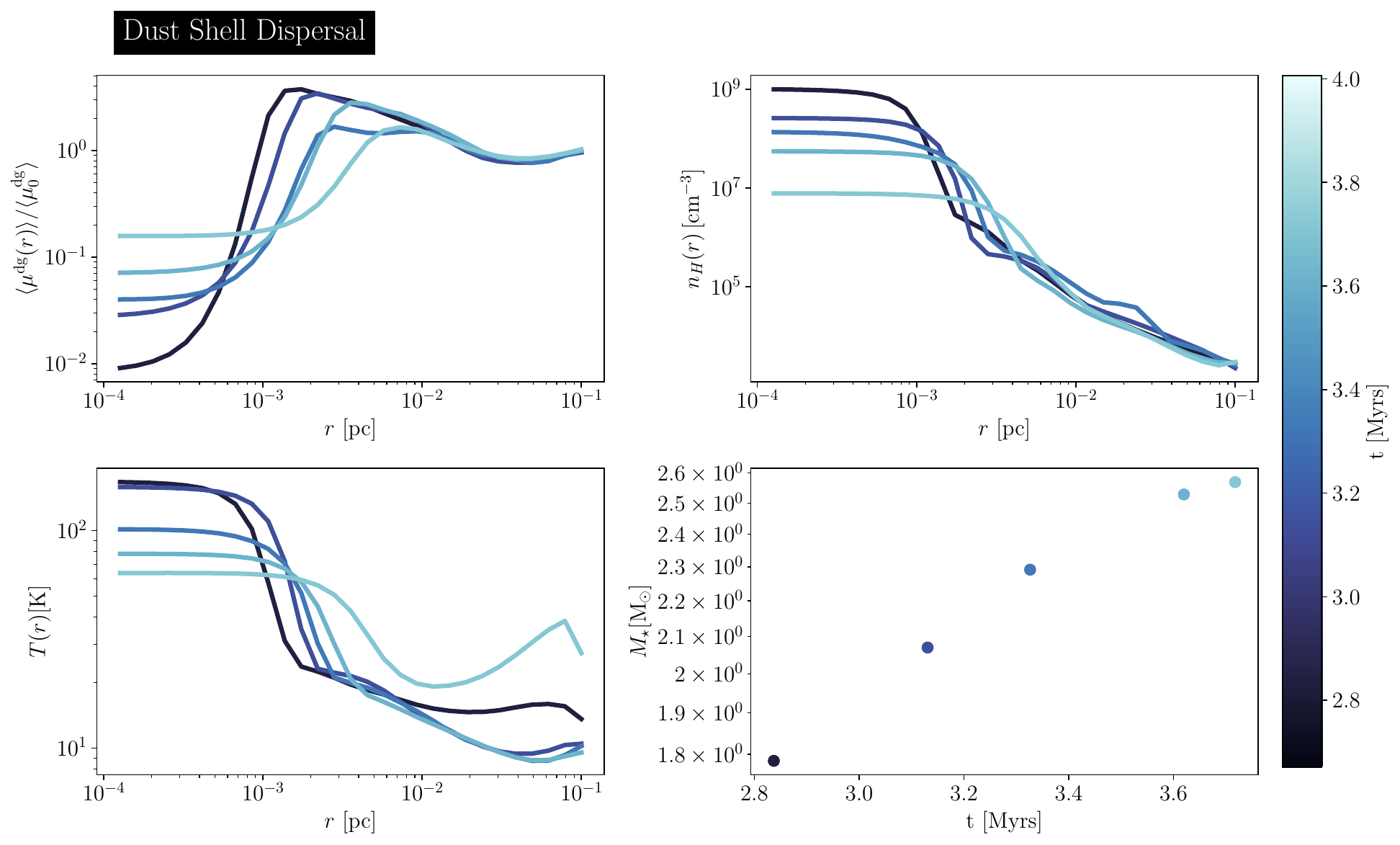}
    \caption{The temporal evolution of environmental properties detailing the creation and subsequent dispersion of a dust-evacuated region around a reference sink particle. We only show specific timesteps that most clearly highlight the evolution of this process. Each line in the plot corresponds to a given timestep and is color-coded accordingly. The upper four plots, in the hot colormap, depict the formation of a dust-evacuated region during a high accretion state, while the lower four, in the cold colormap, show its dispersal during reduced accretion. The properties are evaluated within narrow radial shells. {\bf{Top Left}}: The mean dust-to-gas (DTG) $\langle\dustgas (r)\rangle$ ratio normalized to the initial mean value of the cloud $\langle\dustgas_0 \rangle$. {\bf{Top Right}}: The mean number density of the gas $n_{\rm H}(r)$. {\bf{Bottom Left}}: The mean gas temperature $T(r)$. {\bf{Bottom Right}}: The sink particle's mass evolution. The plots illustrate the in-fall of material towards the sink, raising the central gas density. Ongoing accretion and heightened luminosity increases the gas temperature due to radiation. As gas flows inward, it carries dust closer, but the enhanced luminosity counteracts the motion, propelling dust outward to form a $\sim 10^{-2}$ pc thick dust shell, with a peak DTG ratio of $\langle \dustgas \rangle \sim 5 \langle \dustgas_0 \rangle$. As accretion slows down, the gas density and temperature decrease, while the DTG ratio rises in the inner regions, and the peak shifts outward. The plots illustrate variations in $\dustgas$ around a sink particle induced by radiation pressure, and highlight the development of a dust-evacuated zone encircled by a dust-rich shell.}
    \label{fig:r_profile}
\end{figure*}

In Figure \ref{fig:r_profile}, we present a case study of this phenomenon, focusing on a region around a selected sink particle. We chose this particular candidate after examining several others, as it demonstrates the median behaviour with minimal noise. We also carefully chose specific timesteps to minimise the impact of noise on the data and to clarify the dust evacuation process. Each line in the plot corresponds to a snapshot in time and is colour-coded accordingly. All values are computed as the mean values within narrow radial shells centered around the star. The top four plots represent the state of the system during the initial creation of the dust-evacuated region, while the lower section corresponds to the subsequent dispersal of the dust-evacuated region.

At early times, there is only a minimal reduction in the DTG near the sink particle. For instance, at $10^{-4}$ pc from the sink, the DTG ratio is only half of the average initial value within the cloud. As we move farther from the sink, we observe an accumulation of dust, leading to a DTG ratio of approximately $\dustgas / \dustgas_0 \sim 2$, which gradually approaches the mean value as we reach a distance of about 0.1 parsecs from the sink.

As the sink continues to accrete matter and increase in mass, we observe corresponding changes in the gas environment. The gas number density increases from its peak value of $10^7 \rm cm^{-3}$ to $10^9 \rm cm^{-3}$, and the gas temperature transitions from an initial value of $T\sim 20$K to $T\sim 100 $K. As material accretes onto the sink, the dust is entrained alongside the gas, and the peak in the DTG ratio shifts closer to the sink. Additionally, we observe that the peak DTG ratio increases. This effect is primarily attributed to the gas density dropping more rapidly than the dust density in that specific region. We propose that this effect is driven by radiation pressure forces that induce a net outward motion of the dust, in contrast to the net inflow of gas. This process gives rise to the formation of a dust-evacuated region surrounding the star. As a consequence, a dust shell, approximately $10^{-2}$ pc in thickness, emerges, where $\langle \dustgas_0 \rangle / \langle \dustgas_0 \rangle \sim 5$. In the immediate vicinity of the star, this effect is accompanied by values dropping as low as $\langle \dustgas \rangle / \dustgas_0 \sim 10^{-2}$ in the central region.

In the four bottom plots, we present the same parameters during a phase characterised by a decline in the accretion rate, as indicated by the sink mass-time plot. We observe a reduction in the gas number density and temperature over time, consistent with expectations during a reduced accretion phase. Additionally, we note a gradual increase in the DTG ratio within the innermost regions, attributable to the diminishing gas density near the central region. The reduced gas number density also leads to weaker drag forces experienced by the dust grains, resulting in the outward shift of the peak DTG ratio and the reduction of its maximum value as the dust shell disperses.

\begin{figure*}
    \centering
    \includegraphics[width=0.95\textwidth]{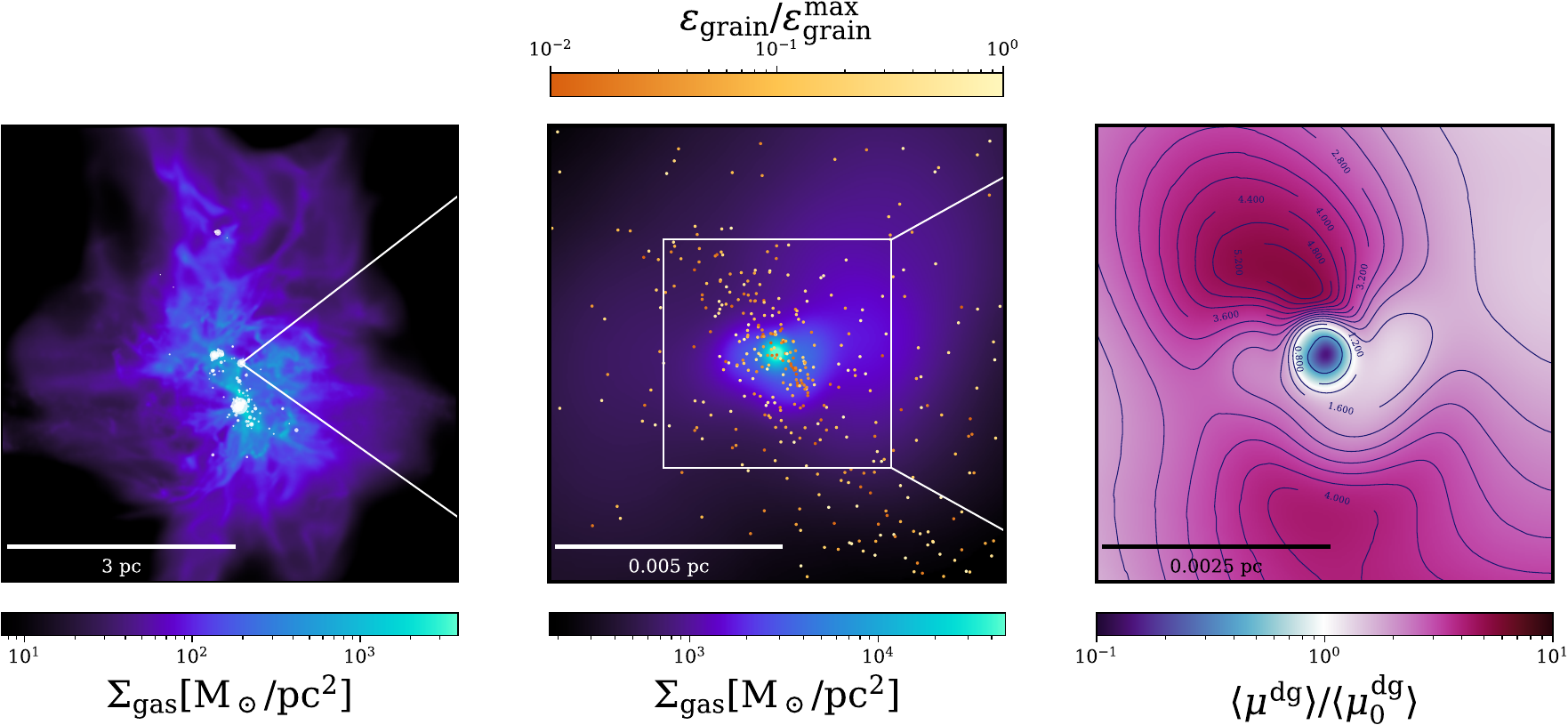}
    \caption{The 2D integrated gas surface density of the m2e3\_0.1\_hires simulation. \textbf{Left}: The gas surface density of the cloud, with white circles representing sink particles, sized according to their mass. \textbf{Middle}: The gas surface density centered around a specific sink particle, with individual dust particles color-coded by grain size. \textbf{Right}: The dust-to-gas mass ratio (DTG) normalized to the mean DTG of the cloud within a 0.01 pc region centered around the sink particle. This figure illustrates the presence of a dust-evacuated zone around a sink particle, featuring a dust pile-up or ``shell'' surrounding the sink.}
    \label{fig:smallscale}
\end{figure*}

While we demonstrated that both dust and gas exhibit similar features on large scales, we will now discuss the differences that arise on smaller scales. In Figure \ref{fig:smallscale}, we transition from the overview of the gas surface density of the entire cloud in our m2e3\_0.1\_hires run ($6 \text{pc} \times 6 \text{pc} \times 6 \text{pc}$) to a reduced volume. Specifically, the middle panel focuses on a smaller volume with dimensions of ($0.01 \text{pc} \times 0.01 \text{pc} \times 0.005 \text{pc}$), presenting the integrated 2D surface density of the gas through a slice in that region. Individual dust particles, color-coded to denote their respective grain sizes, are overlaid on the gas distribution. The right panel zooms in on an even more compact space with dimensions of ($0.005 \text{pc} \times 0.005 \text{pc} \, \times  0.0025\text{pc}$), showcasing the mean DTG ratio within the volume. Both zoomed-in plots are centered around a specifically chosen sink particle. It is worth noting that the selection of this particular sink particle and snapshot is intentional, as they effectively illustrate the dust evacuation and dust pile-up phenomena.

The plot highlights a key observation: on sub-parsec scales, particularly in the proximity of stars, the spatial distribution of dust particles deviates from that of the gas. Further as indicated by the plots in Figure \ref{fig:r_profile}, we observe a dust-suppressed zone near the sink particle followed by dust-rich region, indicative of the presence of a dust pile-up or a dust ``shell'' enveloping these stellar objects. Specifically, the DTG ratio peaks around the dust-depleted region and roughly reverts to the mean beyond the area shown in the plot.

\begin{figure}
    \centering
     \includegraphics[width=0.47\textwidth]{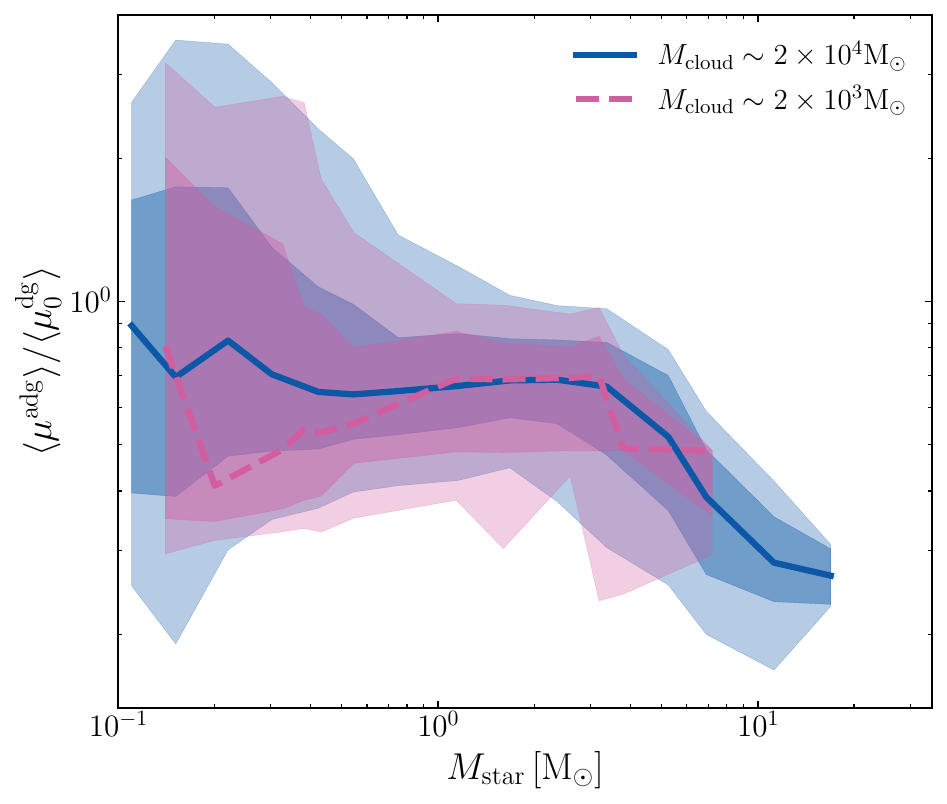}
    \caption{The rolling median of the accreted dust-to-gas (ADG) ratio $\mu^{\rm adg}$ normalized to the initial mean DTG ratio of the cloud $\langle \mu_0^{\rm dg} \rangle$ for clouds with $M_{\rm cloud} \sim 2\times10^3 \rm M_\odot$ and $M_{\rm cloud} \sim 2\times10^4 \rm M_\odot$. We show both clouds simulated at a resolution of $\Delta m_{\rm gas} \sim 10^{-2} \rm M_\odot$. Darkly shaded and lightly shaded regions indicate the interquartile and interdecile percentile ranges, respectively. The relatively consistent trend suggests that, within the mass range investigated here, the cloud mass does not affect the trend.}
    \label{fig:cloud_mass}
\end{figure}
\begin{figure}
    \centering
    \includegraphics[width=0.47\textwidth]{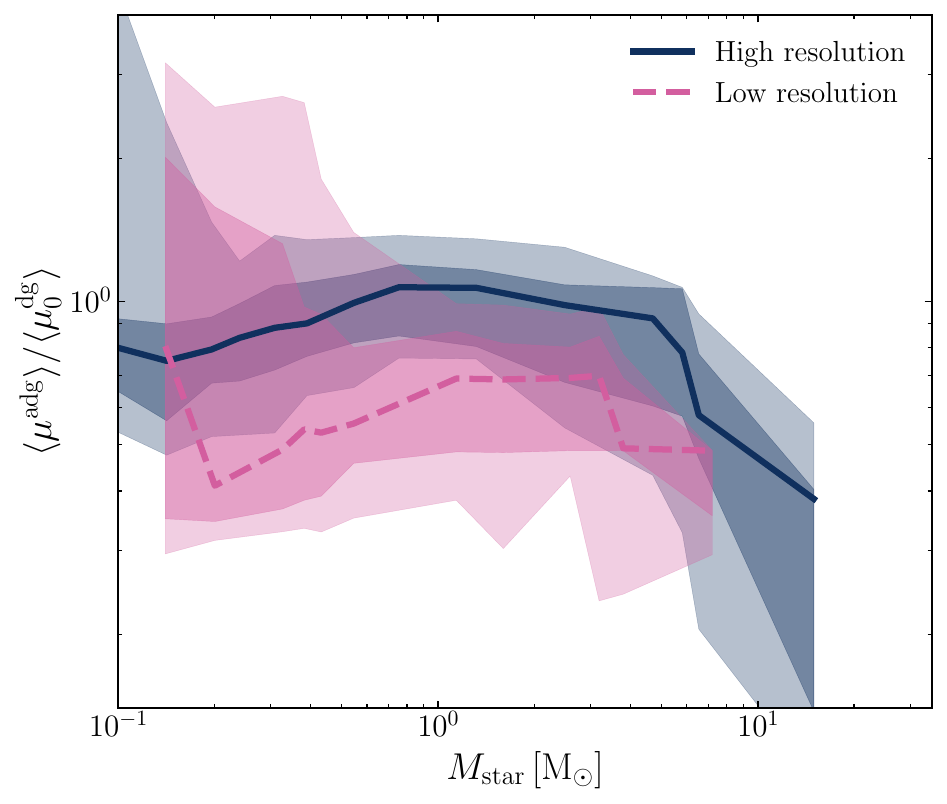}
    \caption{The normalized rolling median of the accreted dust-to-gas ratio ($\mu^{\rm adg}$) in simulations with different resolutions. Shading indicates interquartile and interdecile ranges. Dark blue line shows a $2\times10^3 \rm M_\odot$ cloud at resolutions of $\Delta m_{\rm gas} \sim 10^{-3}$ while the pink line shows the same cloud at lower resolution $\Delta m_{\rm gas}\sim 10^{-2} \rm M_\odot$. Higher resolution reduces scatter, and aligns $\mu^{\rm adg}$ closer to $\langle \mu_0^{\rm dg} \rangle$ before declining at higher sink masses.}
    \label{fig:res}
\end{figure}

\subsection{Effects of Cloud Mass}
We consider how this phenomenon is affected by cloud properties. In Figure \ref{fig:cloud_mass}, we compare the ADG ratio in two distinct clouds, both simulated with our full physics setup. Both clouds are simulated with a common resolution of $\Delta m_{\rm gas} \sim 10^{-2} \rm M_\odot$, with one having a mass of $2\times10^3 \rm M_\odot$ and the other $2\times10^4 \rm M_\odot$ The results obtained reveal a consistent trend in the ADG ratio, which remains unaffected by variations in cloud mass. Across this mass range, the influence of radiation pressure, inherently a localised phenomenon, exerts comparable effects across diverse cloud masses. Nevertheless, although this phenomenon primarily impacts the ADG ratio of a specific star, its influence could extend to larger radii. As more massive stars form, their high luminosities could also lead to a reduction in dust content for neighbouring stars. 

While the primary emphasis of this work centers on reduced ADG ratios for high-mass stars, our simulations also unveil a subset of low-mass stars that are rich in dust content. Specifically, within the subset of solar and subsolar mass stars, we observe instances of heightened ADG ratios. However, we refrain from attributing this phenomenon to a single cause. Instead, we postulate that it likely stems from the intricate interplay between turbulent motions and radiation pressure emanating from neighboring stars, leading to the redistribution of dust. Hence, stars emerging from regions with dust over-densities are prone to experiencing increased levels of dust accretion.

\subsection{Effects of Simulation Resolution}
\label{sec:resol}

In Figure \ref{fig:res}, we present a comparative analysis of the ADG ratios within stellar populations formed in a cloud with a mass of $M_{\rm cloud} \sim 2 \times 10^{3}\rm M_\odot$, utilising simulations at different resolutions $\Delta m_{\rm gas} \sim 10^{-3}$ and $\Delta m_{\rm gas}\sim 10^{-2} \rm M_\odot$). As anticipated, our observations indicate that increasing the simulation resolution decreases the scatter in the distribution. In addition, for the higher resolution cloud, the ADG ratio tends to converge to $\dustgas_0$ for relatively low-mass stars (those with masses less than a few solar masses). This aligns with the expectation that stars with lower luminosities lack the necessary conditions to evacuate dust, and therefore would have ADG ratios that mirror the average DTG ratio within the cloud. These improvements can be attributed to the enhanced sampling capabilities, allowing for a more precise resolution of accretion and sink formation.

However, as shown in the plot, our findings are sensitive to resolution. This observation is unsurprising given the inherent challenges in accurately modelling accretion around luminous stars \citep{krumholz2009formation, krumholz:2012.rad.pressure.rt.instab, rosen2016unstable}. Achieving an accurate depiction of relative dust-to-gas accretion onto a star, accounting for associated radiation effects, requires significantly higher resolutions than currently feasible with our simulations.

To illustrate, considering a 1 solar mass star that would have accreted approximately $\sim 0.01 M_\odot$ of dust mass, assuming $\dustgas \sim 10^{-2}$, at the $10^{-2}$ resolution (bearing in mind that the dust is up-sampled by a ratio of 4 times the gas resolution), this roughly corresponds accretion of $\sim$ 400 dust particles. While this number of particles is sufficient to resolve the phenomenon, noise may still depend on resolution, potentially limiting our ability to capture more complex dynamics that occur on smaller scales. Consequently, we do not necessarily anticipate convergent results at our low resolutions, particularly at low sink masses. However, this approach still provides valuable insights when comparing simulations conducted at the same resolution. We hope that our results motivate detailed zoom-in simulations of dusty accretion around protostellar objects and onto massive stars, to provide a more detailed study of the dust evacuation phenomenon.

To evaluate the effects of resolution on our findings, we conducted a series of idealized tests of singular \citet{shu_1977_isothermal_collapse} collapse scenarios at resolutions of $10^{-2}$, $10^{-3}$, and $10^{-4} \rm M_\odot$. These tests track the problem hydrodynamically, allowing sink particle formation with subsequent accretion and radiation. To simplify our analysis, we exclude the effects of magnetic fields and protostellar jets and stellar winds. We find that the observed phenomena qualitatively persist across different resolutions; however, convergence is not achieved. In particular, neither the spatial extent of the dust evacuation zone nor the precise magnitude of the stellar ADG ratio converge with increasing resolution.

Considering the complexities of radiation-limited accretion, it remains uncertain whether the system, as modeled in our simulations, should converge within our resolution range. Higher resolution simulations capture finer-scale structures, unveiling more intricate dynamics around the sink which thereby altering the magnitude and scale of dust evacuation. Additionally, the mechanisms of dust and gas accretion, whether via spherical collapse or disk accretion, and their dynamics are resolution-dependent and influenced by the specific physics in our simulations.

\subsection{Effects of Altered Grain Size Distributions}
\label{sec:grainsize}
\begin{figure}
    \centering
    \includegraphics[width=0.47\textwidth]{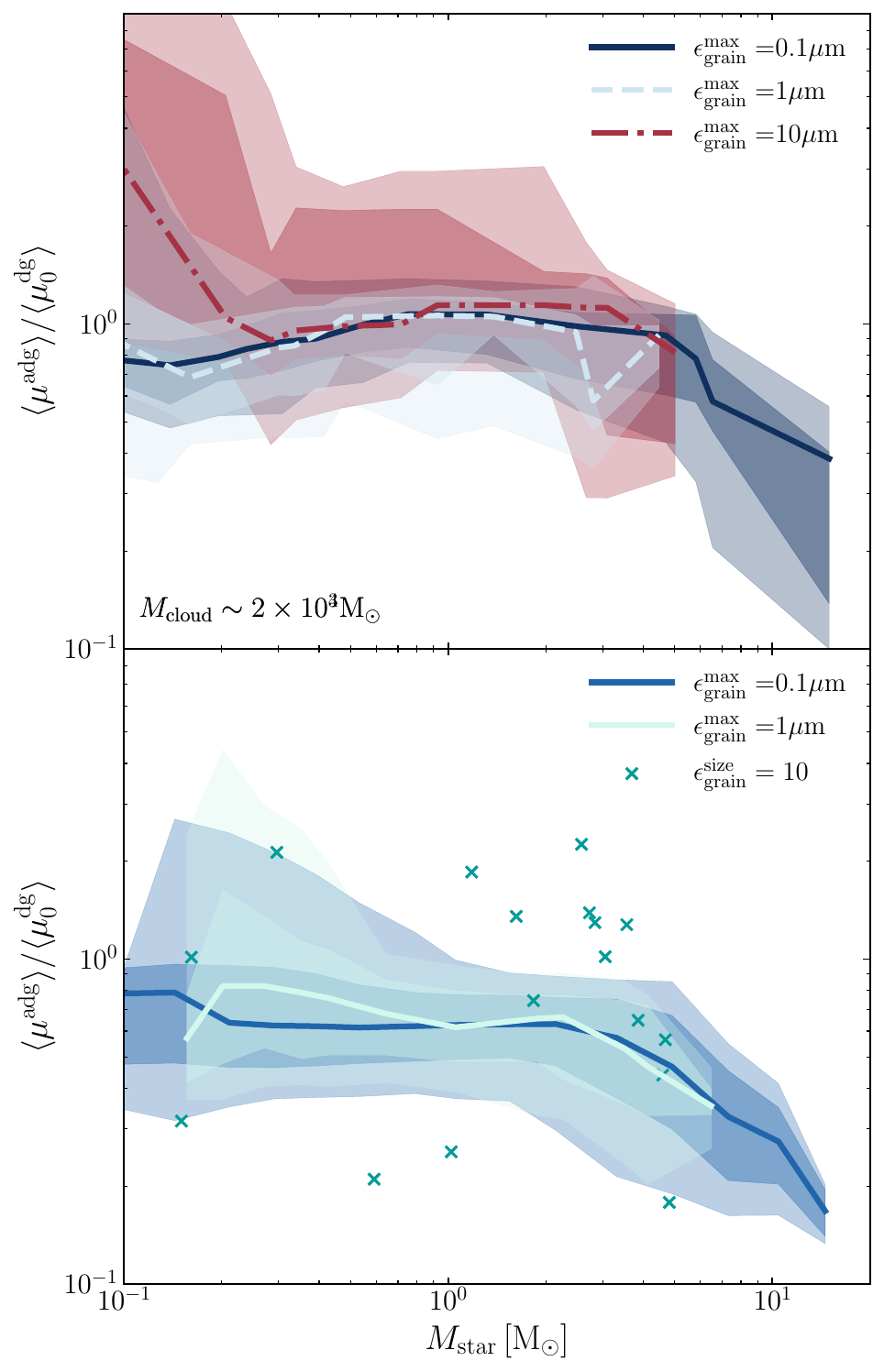}
    \caption{The rolling median of the accreted dust-to-gas (ADG) ratio ($\mu^{\rm adg}$) for sink particles formed within a cloud of initial mass $M_{\rm cloud} = 2 \times 10^3 \rm M_\odot$ (top), $M_{\rm cloud} = 2 \times 10^4 \rm M_\odot$ (bottom) with different initial grain-size distributions ($\grainsizemax=0.1 \rm \mu m$, $\grainsizemax=1 \rm \mu m$, and $\grainsizemax=10 \rm \mu m$).  The darkly shaded and lightly shaded regions represent the interquartile and interdecile percentile ranges, respectively. The m2e4\_10 run exhibits notable scatter and a low count of formed sink particles. To account for this, we scatter the ADG ratio for each sink particle. Refer to Section \ref{sec:resol} for a discussion on potential drivers of this scatter. Changes in the grain size do not drive significant deviations from the reduced $\mu^{\rm adg}$ trend for high mass stars, but larger grain sizes lead to fewer high-mass stars due to reduced star formation efficiency.}
    \label{fig:grainsize}
\end{figure}

An additional variable to consider is the size of the dust grains. At extremely small grain sizes, the grains should closely trace the dynamics of the gas, whereas at larger grain sizes, the grains will be entirely decoupled from gas dynamics. In Figure \ref{fig:grainsize}, we explore the impact of various plausible grain sizes within the cloud environment, incorporating grains with maximum sizes of $\epsilon_{\rm grain}^{\rm max} \sim 10 \rm \mu m$ and $\epsilon_{\rm grain}^{\rm max} \sim 1 \rm \mu m$, alongside our baseline assumption of $\epsilon_{\rm grain}^{\rm max} \sim 0.1 \rm \mu m$. This is conducted while ensuring a fixed total dust mass within the simulation and maintaining a dynamic range of $\sim 100$ in grain sizes. We show the results for our higher resolution $M_{\rm cloud} \sim 2 \times 10^3 \rm M_{\odot}$ simulations, in addition to a $M_{\rm cloud} \sim 2 \times 10^4 \rm M_{\odot}$ cloud at lower resolution. While in principle, grain size could influence our results, our findings for these more reasonable sizes are consistent with Equation \ref{eq:rad}, which demonstrates no grain-size dependence. It is worth noting that the grain size does have an impact on the total number of sink particles that form, with larger grains leading to a reduced overall sink count. This effect and related changes will be studied in detail in Soliman \& Hopkins (in prep.).

However, we begin to see some deviations from the reported trend for our largest grain simulations. Note that we present individual $\mu^{\rm adg}$ values for each sink particle in the simulation with $M_{\rm cloud} \sim 2 \times 10^4 \rm M_{\odot}$ and $\grainsizemax = 10 \rm \mu m$, opting against plotting the median due to a limited number of sink particles and a substantial dispersion in their $\mu^{\rm adg}$. The dispersion is likely influenced by several factors, including the relatively lower resolution. However, we posit that the increased grain size inherently contributes to a higher level of scatter. Recall that these simulation assume a fixed a grain size spectrum, where individual grain particles maintain a fixed size throughout the simulation. Consequently, as grain size increases, the individual grain count decreases, while each grain becomes more massive. This suggests that whether a particle undergoes accretion or expulsion has a more pronounced impact on the overall dust mass accreted by a sink, especially when compared to its smaller size/mass grain counterpart. Further, Equation \ref{eq:rad} assumes that $\langle Q \rangle_{\rm ext}$ is proportional to $2 \pi \grainsize/\lambda$, where $\lambda$ denotes the radiation wavelength. However, for grains larger than $2 \pi/\lambda$, $\langle Q \rangle_{\rm ext}$ approaches 1. As a result, $\driftvelmag \propto \grainsize^{-1}$, introducing an inverse dependence on grain size into Equation \ref{eq:rad}. Considering the case of clouds with $\grainsizemax = 10 \rm \mu m$, all grains fall within this regime for radiation with $\lambda \lesssim 600 \rm \mu m$, and larger grains fall into this regime for even shorter wavelengths. This wavelength range corresponds to the optical/UV range, where stars—especially more massive and hotter ones—typically emit peak radiation. Therefore, we would expect to see increases in $\mu^{\rm adg}$ for such massive grains, although such large grains are not expected to be abundant in most GMCs.

In reality, dust grains undergo processes such as coagulation, accretion, sputtering, photodestruction, and shattering which result in both changes in the grain size distribution and the overall dust mass. These mechanisms, currently not captured in our simulations, could impact the interpretation of our results. For instance, if dust sublimation proceeds efficiently and the elements comprising the grains become well-mixed in the gas phase near the stars, they may be incorporated into the stars, potentially erasing the effects of dust evacuation on stellar abundances. However, the dust-evacuated zones around the stars would still persist or undergo further evacuation. We acknowledge these limitations and plan to address them in future work.

\begin{figure*}
    \centering    
    \includegraphics[width=0.95\textwidth]{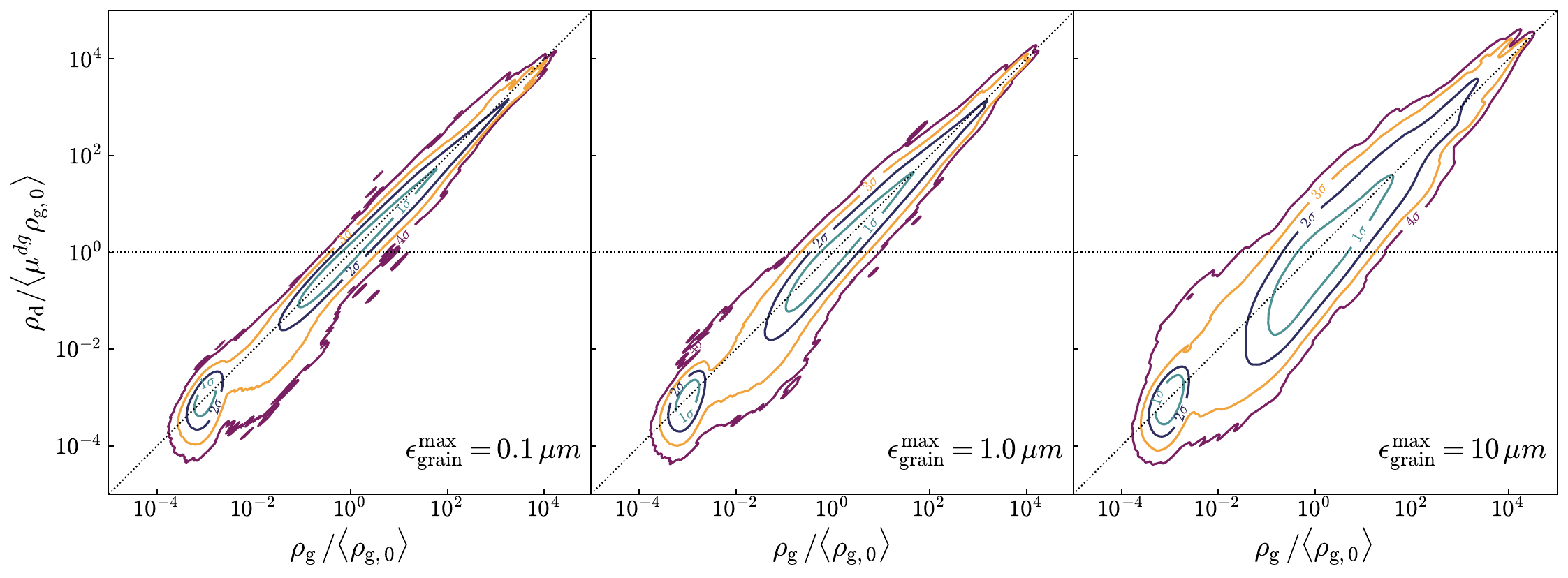}
    \caption{Bivariate distribution of the local 3D gas density ($\rho_g$) and dust density ($\rho_d$), weighted by dust mass, illustrating the probability distribution around a grain at the resolution scale of approximately $10^{-2}$ pc. From left to right, we show the distribution for a run with $\grainsizemax =0.1, 1, 10 \, \rm \mu m$ respectively for our $M_{\rm cloud} \sim 2\times 10^3 \rm M_\odot$ at $\Delta m_{\rm gas} \sim 10^{-3} \rm M_\odot$. We show the $1-\sigma$ (green), $2-\sigma$ (navy), $3-\sigma$ (orange), $4-\sigma$ (plum) contours. The diagonal dotted lines represent perfect dust-gas coupling ($\rho_d = \mu^{\rm dg} \rho_g$), while the horizontal line denotes uniform dust density. The distribution becomes progressively broader indicative of weaker coupling for larger grain sizes. Specifically, we find that the Pearson correlation coefficient is R = 0.994 for $\grainsizemax = 0.1 \rm \mu m$, R = 0.990 for $\grainsizemax = 1 \rm \mu m$, and R = 0.967 for $\grainsizemax = 10  \rm \mu m$.}
    \label{fig:coupling}
\end{figure*}

In Figure \ref{fig:coupling}, we present the bivariate distribution showing the correlation between the local 3D gas density ($\rho_g$) and the dust density ($\rho_d$). This distribution is weighted by the dust mass and computed at a spatial resolution of approximately $10^{-2}$ pc for the different grain size runs ($\epsilon^{\rm max}_{\rm grain} = 0.1, 1, 10 , \mu \text{m}$) at a time of approximately $2 \, t_{\rm dyn}$. For the smallest grains, the distribution exhibits a relatively narrow distribution, primarily centered around the theoretically predicted perfect coupling line $\rho_d = \mu^{\rm dg} \rho_g$. However, as the grain size increases, the distribution broadens, signifying a decreased level of coupling between the gas and dust components. Nevertheless, on average, the fluid remains sufficiently well-coupled to prevent large fluctuations in the DTG ratio.

\section{Discussion}
\label{sec:dis}
\subsection{Implications}

Our findings, which reveal a diminished accretion of dust onto massive stars under the influence of radiation pressure, bear notable implications for the stellar abundances of CNO, Mg, Fe, and Si, elements known to preferentially deplete onto dust grains. This phenomenon was previously suggested in the literature as a potential explanation for the anomalous abundance ratios observed in nearby stars \citep{melendez2009peculiar, gustafsson2018dust1, gustafsson2018dust2, mathews1967dynamic, cochran1977development}. A distinctive feature of this model is its localized effect, with the evacuation zone extending up to $\sim 10^{-3}$ pc or a few hundred AUs. Consequently, this preferential accretion mechanism may contribute to elucidating the anomalous abundance ratios observed in specific nearby stars and the variations in abundance ratios among stellar pairs \citep[e.g.,]{maia2014high, nissen2017high, ramirez2015dissimilar, biazzo2015gaps, teske2016curious, teske2016magellan, saffe2017signatures, oh2018kronos}.

Further, the observed decrease in dust content has significant implications for the gas dynamics and thermochemistry near the star. Lower dust content, associated with diminished opacities and cooling rates, would accelerate the expansion rate of HII regions \citep{ali2021growth}. It would likely result in a larger ionized volume, owing to the reduced UV absorption by dust.

Moreover, protoplanetary disks would also potentially affected by this phenomenon. The DTG ratio plays a crucial role in determining the efficiency of dust radial drift, influencing the structural evolution of the disk \citep{toci2021secular}. Moreover, as dust constitutes the fundamental building block of planets, diminished dust content is likely to influence the types of planets that can form and their compositions.

\subsection{Caveats}
\label{sec:caveats}
The current study has inherent limitations that warrant careful consideration. A major constraint is the absence of a model for dust evolution; our approach relies on assuming a constant grain size distribution and a constant total dust mass throughout the simulation. Consequently, should the grain size distribution undergo significant changes, specific conclusions within our study may require reevaluation. In particular, if dust grains experience substantial destruction, whether through sublimation, shattering and/or sputtering, evacuated regions would likely persist or even intensify. However, if the elements released from the grains transition to the gas phase and continue to accrete onto stars, spatial variations in dust distribution might not necessarily indicate corresponding variations in stellar abundances.

Furthermore, our fiducial grain size distribution incorporates nanometer-sized grains. The existence of such small grains in these conditions is disputed, given the potential for coagulation or destruction due to the conditions in molecular clouds. Additionally, it is uncertain whether such grains can be modeled aerodynamically and whether they would adhere to the charge and mass scalings we assume.Specifically, the similarity in size of these small grains to the large background molecules they interact with makes it unclear if a simple Epstein drag model can accurately capture their dynamics, as it does not account for effects such as deformation and non-spherical geometries. Additionally, at these scales, effects such as charge quantization become significant. However, it is important to note that the results presented here predominantly depend on the grains holding the most mass, which, under an MRN spectrum, are the largest grains. Nevertheless, these considerations prompt us to explore different ranges of grain size in Section \ref{sec:grainsize} to assess how varying grain properties might influence our results.

An additional constraint in our study is the resolution limit, which is discussed in more detail in Section \ref{sec:resol}. To ensure a statistically robust sample size for identifying the phenomenon of dust evacuation, we conducted global star formation simulations. However, the most interesting behavior occurs at the resolution limit of our study. Achieving detailed predictions for the implications of dust dynamics on stellar metallicities, protostellar envelope properties, and planet formation requires more intricate small-scale physics. Addressing these complexities realistically demands a substantial increase in resolution, which we plan to explore in future work.

\section{Conclusions}
\label{sec:conc}

This study introduces a set of RDMHD simulations of star forming GMCs as part of the {\small STARFORGE} project. These simulations encompass a detailed representation of individual star formation, accretion, and feedback mechanisms, while also explicitly considering the influence of the dynamics of dust grains. Our investigation focuses on the implications of these dynamics, specifically radiation-dust interactions, on the emergent properties of stellar populations.

Through our analysis, we find that when stars surpass a critical mass threshold ($\sim 2 M_{\odot}$), their luminosity exerts sufficient radiation pressure on neighbouring dust grains, ultimately resulting in their expulsion from the star's accretion radius. This drives the formation of a dust-evacuated region of size $\sim 100$ AU. Consequently, this process results in a mass-dependent adjustment in the ADG mass ratio incorporated into these stars via the accretion process. Furthermore, we investigate the potential implications of varying cloud mass, grain sizes, and other physical parameters such as decoupling the grains from magnetic fields and deactivating stellar winds and jets. However, our findings indicated that these variations had negligible effects within the parameter space of our study. 

 In summary, our findings shed light on the interplay between radiation, dust dynamics, and star formation, offering valuable insights into the complex processes that shape stars, their environments, and compositions within molecular clouds.

\vspace{1em}

\noindent Support for for NS and PFH was provided by NSF Research Grants 1911233, 20009234, 2108318, NSF CAREER grant 1455342, NASA grants 80NSSC18K0562, HST-AR-15800. Support for MYG was provided by NASA through the NASA Hubble Fellowship grant \#HST-HF2-51479 awarded  by  the  Space  Telescope  Science  Institute,  which  is  operated  by  the   Association  of  Universities  for  Research  in  Astronomy,  Inc.,  for  NASA,  under  contract NAS5-26555. Numerical calculations were run on the TACC compute cluster ``Frontera,'' allocations AST21010, AST20016, and AST21002 supported by the NSF and TACC, and NASA HEC SMD-16-7592.  This research is part of the Frontera computing project at the Texas Advanced Computing Center. Frontera is made possible by National Science Foundation award OAC-1818253.

\datastatement{The data supporting this article are available on reasonable request to the corresponding author. }

\software{\href{https://matplotlib.org/}{\fontfamily{cmtt} \selectfont matplotlib} \citep{Hunter:2007}, \href{https://numpy.org/}{\fontfamily{cmtt} \selectfont NumPy} \citep{harris2020array}, \href{https://scipy.org/}{ \fontfamily{cmtt} \selectfont SciPy} \citep{2020SciPy-NMeth}, \href{https://cmasher.readthedocs.io/index.html}{\fontfamily{cmtt} \selectfont CMasher}\citep{cmasher}.}
\bibliography{sample631}{}
\bibliographystyle{aasjournal}



\end{document}